\documentclass[apj]{emulateapj}
\usepackage{times}
\usepackage{subfigure}
\usepackage{graphicx}
\usepackage{setspace}
\usepackage{amssymb}
\usepackage{multirow}
\usepackage{color}
\usepackage{wrapfig}
\usepackage{float}
\usepackage{lineno}
\usepackage{amsmath}
\usepackage{url}

\newif\ifAMStwofonts
\AMStwofontstrue

\makeatletter

\newcommand{\Rmnum}[1]{\expandafter\@slowromancap\romannumeral #1@}

\newcommand{\degree}{^{\circ}}
\newcommand{\sech}{\,\mathrm{sech}}
\newcommand{\myemail}{lujig@pku.edu.cn}

\makeatother

\shorttitle{Pulse Profiles of PSR~B1133$+$16}

\shortauthors{Lu et al.}

\begin{document}

\title{Multi-frequency Radio Profiles of PSR B1133$+$16: radiation location and particle energy}

 \author{
J.~G.~Lu\altaffilmark{1,2},~
Y.~J.~Du\altaffilmark{3},~
L.~F.~Hao\altaffilmark{4},~
Z.~Yan\altaffilmark{5},~
Z.~Y.~Liu\altaffilmark{6},~
K.~J.~Lee\altaffilmark{7},~
G.~J.~Qiao\altaffilmark{2},~
L.~H.~Shang\altaffilmark{8},~
M.~Wang\altaffilmark{4},~
R.~X.~Xu\altaffilmark{1,2,7},~
Y.~L.~Yue\altaffilmark{8},~
and
Q.~J.~Zhi\altaffilmark{9}
}

 \altaffiltext{1}{State Key Laboratory of Nuclear Science and Technology, Peking University, Beijing 100871, China; \myemail}

 \altaffiltext{2}{School of Physics, Peking University, Beijing 100871, China}

 \altaffiltext{3}{Qian Xuesen Laboratory of Space Technology,  Beijing 100094, China}

 \altaffiltext{4}{Yunnan Astronomical Observatory, Chinese Academy of Sciences, Kunming 650011, China}

 \altaffiltext{5}{Shanghai Astronomical Observatory, Chinese Academy of Sciences, Shanghai 200030, China}

 \altaffiltext{6}{Xinjiang Astronomical Observatory, Chinese Academy of Sciences, Urumqi, Xinjiang 830011, China}

 \altaffiltext{7}{Kavli Institute for Astronomy and Astrophysics, Peking University, Beijing 100871, China}

 \altaffiltext{8}{National Astronomical Observatories, Chinese Academy of Sciences, Beijing 100012, China}

 \altaffiltext{9}{School of Physics and Electronic Science, Guizhou Normal University, and NAOC-GZNU Center for Astronomy Research and Education, Guiyang, Guizhou 550001, China}


\begin{abstract}
The pulse profile of PSR B1133$+$16 is usually regarded as a conal double structure. However, its multi-frequency profiles cannot simply be fitted with two Gaussian functions, and a third component is always needed to fit the bridge region (between two peaks). This would introduce additional, redundant parameters. In this paper, through a comparison of five fitting functions (Gaussian, von Mises, hyperbolic secant, square hyperbolic secant, and Lorentz), it is found that the square hyperbolic secant function can best reproduce the profile, yielding an improved fit. Moreover, a symmetric 2D radiation beam function, instead of a simple 1D Gaussian function, is used to fit the profile. Each profile with either well-resolved or not-so-well-resolved peaks could be fitted adequately using this beam function, and the bridge emission between the two peaks does not need to be a new component. Adopting inclination and impact angles based on polarization measurements, the opening angle ($\theta_{\mu 0}$) of the radiation beam in a certain frequency band is derived from beam-function fitting. The corresponding radiation altitudes are then calculated. Based on multi-frequency profiles, we also computed the Lorentz factors of the particles and their dispersion at those locations in both the curvature-radiation and inverse-Compton-scattering models. We found that the Lorentz factors of the particles decrease rapidly as the radiation altitude increases. Besides, the radiation prefers to be generated in an annular region rather than the core region, and this needs further validation.
\end{abstract}



\section{Introduction}

PSR B1133$+$16 is one of the nearest pulsars, located at a
distance of¡¡
$d=350^{+20}_{-20}~\mathrm{pc}$~\citep{bris02}. It has a period of
$P=1.1879~\mathrm{s}$~\citep{hobb04} and is characterized by a
dispersion measure of
$\mathrm{DM}=4.8451~\mathrm{cm^{-3}~pc}$~\citep{hass12}.
Its pulse profiles have previously been analyzed to map the
radius-frequency relations~\citep{cord78,thor91,xilo96,mitr02}.
Because of the long period and small dispersion measure of PSR
  B1133$+$16, the impact on its profile of the effects of both
  dispersion and scattering is small. This renders its profile highly
suitable for analysis of the radio radiation beam.
The profiles of PSR B1133$+$16 are usually treated as ``well-resolved
conal-double'' profiles~\citep{rank83}, and it is therefore
natural to fit the profiles with a two-component model.
However, the integrated profiles cannot be fitted satisfactorily in
the bridge region between the peaks, and hence an additional component
had to be added to optimize the fit results~\citep{hass12}.
In fact, an obvious bulge appears in simulated low-frequency
profiles when using three components to simulate the
profiles~\citep{hass12}, which, however, is not present in real
profiles~\citep{phil92,karu11}.
Based on the derived best-fitting parameters, the different radiation
models and the location of the radiation could be verified and validated.

The radiation mechanism of pulsar radio emission is still not
well-understood.
Nevertheless, a variety of models have been proposed, and certain
pulse profiles are predicted by different models.
In the curvature-radiation (CR) model, the radiation beam is always
regarded as a single conal component. Consequently, the pulse profile
should be composed of either a single peak or double
peaks~\citep{rude75,wang13}.
The nature of the core emission in the CR model has also been
discussed~\citep{gil90,besk12}. It could explain some
pulsars' pulse profiles.
On the other hand, the inverse-Compton-scattering (ICS) model includes
one core and two conal components, and thus the number of pulse peaks
could range between one and five~\citep{qiao98,qiao01}. This could be
consistent with the observational evidence~\citep{rank83,hank86}.
Based on observations of the pulse profiles, a beam composed of one
core and two conal components could be
plausible~\citep{rank83}, but the radio radiation mechanism
of pulsars has not yet been determined unequivocally.
The radio radiation location of pulsars is yet another
puzzle.
Even if just the so-called inner-gap model is considered, one needs to
distinguish between the core gap model~\citep{rude75} and the annular
gap model~\citep{qiao04a}.

In this paper, a conal double model is used to fit the PSR B1133$+$16
profiles at several frequencies.
In Section 2, the observation and our data reduction approach are
introduced.
The pulse profiles are fitted using the conal double model in Section
3.
The energy of particles in the magnetosphere is analyzed by
  considering the radiation mechanism, as discussed in Section 4.
Some remaining issues are discussed in Section 5, and our conclusions
are presented in Section 6.

\section{Observation and Data Reduction}

\begin{table}[!htb]
\begin{center}
\caption{Observation Information.
  \label{obs}}
\begin{tabular}{ll}
\hline
\hline
\multicolumn{2}{c}{2256 MHz data}  \\
\hline
Telescope                                   &      Kunming 40~m radio telescope \\
Number of pulse phase bins                  &      512  \\
Center frequency (MHz)                      &      2256  \\
Bandwidth (MHz)                             &      251.5  \\
System temperature (K)    &    80~\citep{hao10}   \\
SEFD$^a$ (Jy)      &       350~\citep{hao10}    \\
Observation duration (s)                    &      38983  \\
Start UT                                    &      07:17:02  \\
Start MJD                                   &      56802  \\
Start second                                &      26240  \\
\hline
\hline
\multicolumn{2}{c}{8600 MHz data}  \\
\hline
Telescope                                   &      Tianma 65~m radio telescope  \\
Number of pulse phase bins                  &      1024  \\
Center frequency (MHz)                      &      8600  \\
Bandwidth (MHz)                             &      800  \\
System temperature (K)    &    35   \\
SEFD (Jy)      &       50~\citep{yan15}    \\
Observation duration (s)                    &      5354  \\
Start UT                                    &      11:44:49  \\
Start MJD                                   &      56836  \\
Start second                                &      42288  \\
\hline
\end{tabular}
\end{center}
\footnotesize{$^a$ System equivalent flux density.}
\end{table}

Observations at a central frequency of 2256 MHz were made with
the Kunming 40~m Telescope, located at Yunnan Astronomical
Observatory, Chinese Academy of Sciences (CAS).
This telescope is located in Yunnan Province, China, at longitude
102$^\circ$.8 E and latitude +25$^\circ$.0 N~\citep{hao10}.
The signal was folded using the Pulsar Digital Filter Bank 3 (PDFB3)
with parameters provided by PSRCAT\footnote{see
  http://www.atnf.csiro.au/research/pulsar/psrcat/} (version 1.51).
The observations are affected by radio-frequency interference
  (RFI) due to the wireless-fidelity (Wi-Fi) network. This noise was
removed manually with the software package
PSRCHIVE\footnote{http://psrchive.sourceforge.net/}.
Observations at a central frequency of 8600 MHz were obtained with the
Tianma 65~m Telescope, located at longitude 121$^\circ$.1 E and
  latitude +30$^\circ$.9 N. The telescope is located at the Shanghai
  Astronomical Observatory, CAS.
The 8600 MHz signal was folded using the Digital Backend System
(DIBAS).
The relevant observational information is included in Table~\ref{obs}.
The raw data at these two frequencies were added to profiles with PSRCHIVE, respectively.
The 317~MHz data were obtained with the Giant Metrewave Radio
Telescope (GMRT).

\begin{figure}[!t]
\centering
\includegraphics[width=0.4\textwidth]{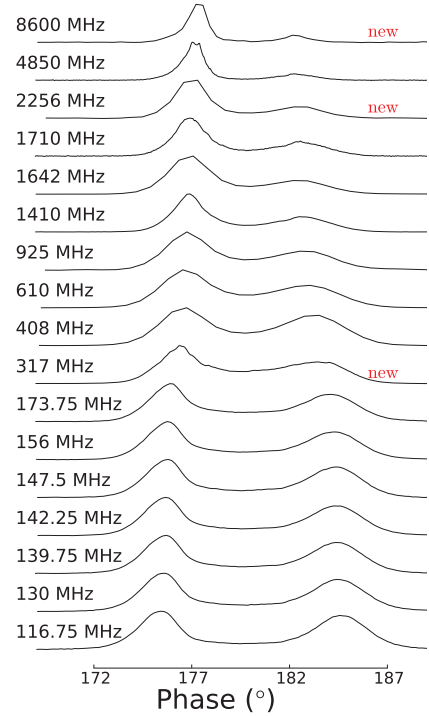}
\caption{Multi-frequency profile of PSR B1133$+$16. All profiles have
  been normalized by their maximum flux densities and are aligned at
  phase $\phi_0$, which was obtained by fitting the profile to
  Eq.~\ref{eq1}.
\label{profile}}
\end{figure}

Apart from the 317, 2256, and 8600 MHz data, other
observations with central frequencies of 116.75, 130, 139.75,
142.25, 147.5, 156, 173.75 MHz~\citep{karu11},
408, 610, 925, 1642 MHz~\citep{goul98}, 1410, 1710, 
and 4850 MHz~\citep{hoen97} have also been adopted here. The data
for the final seven frequencies were downloaded from the European
Pulsar Network (EPN) Data
Archive\footnote{http://www.naic.edu/~pulsar/data/archive.html}.
The multi-frequency profile of PSR B1133$+$16 is shown in
Fig.~\ref{profile}. All profiles have beem normalized by their maximum
flux densities and are aligned at phase $\phi_0$, which was obtained
by fitting the profile to Eq.~\ref{eq1}.
The profile of PSR B1133$+$16 is well resolved, exhibiting a
conal double structure.

\section{Pulse Profile Component Analysis of PSR B1133$+$16}

\subsection{Numerical Experiment Assessing the Profiles of a Single Conal Component}

\begin{figure}[!htb]
\centering
\includegraphics[width=0.5\textwidth]{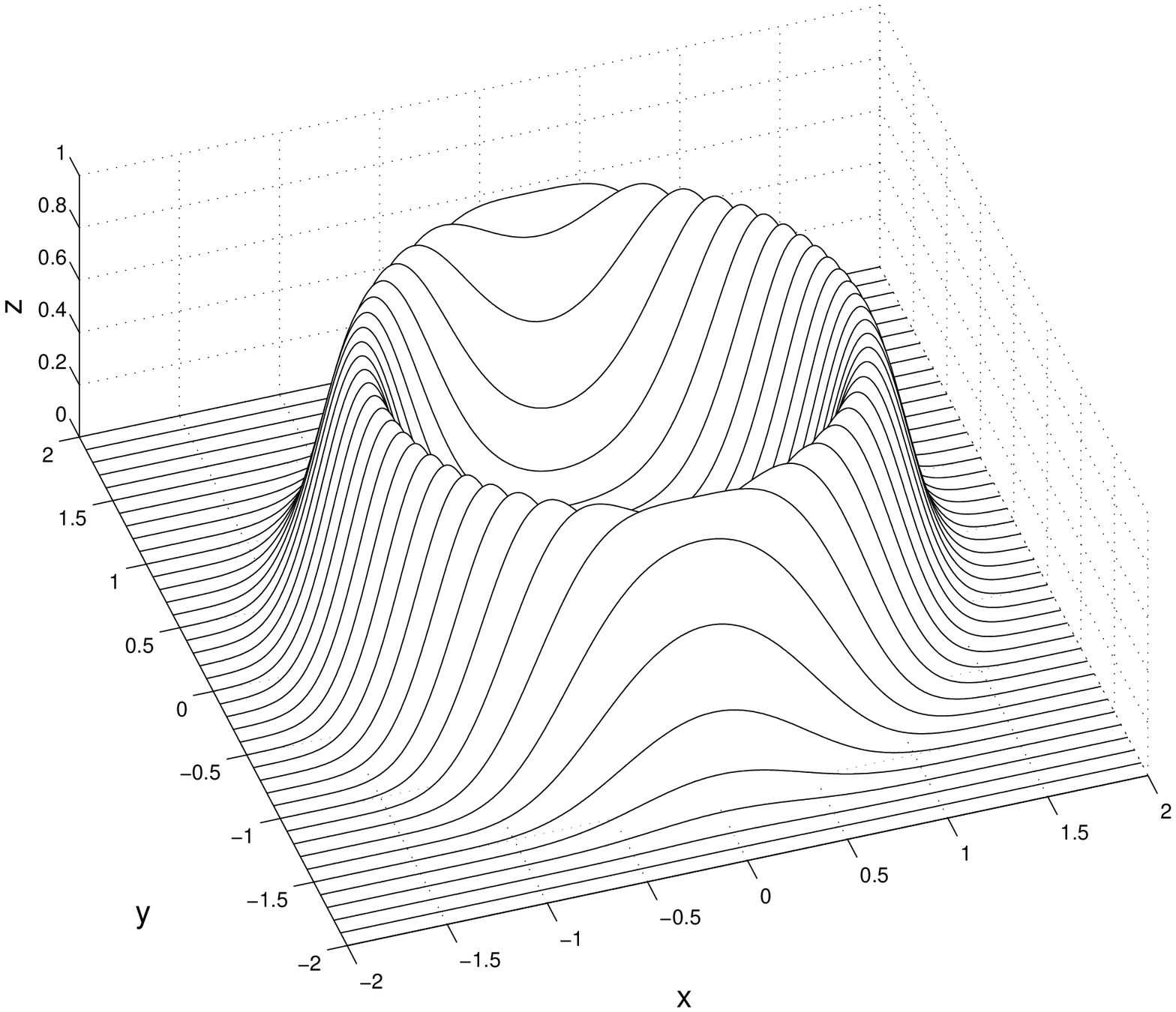}
\includegraphics[width=0.5\textwidth]{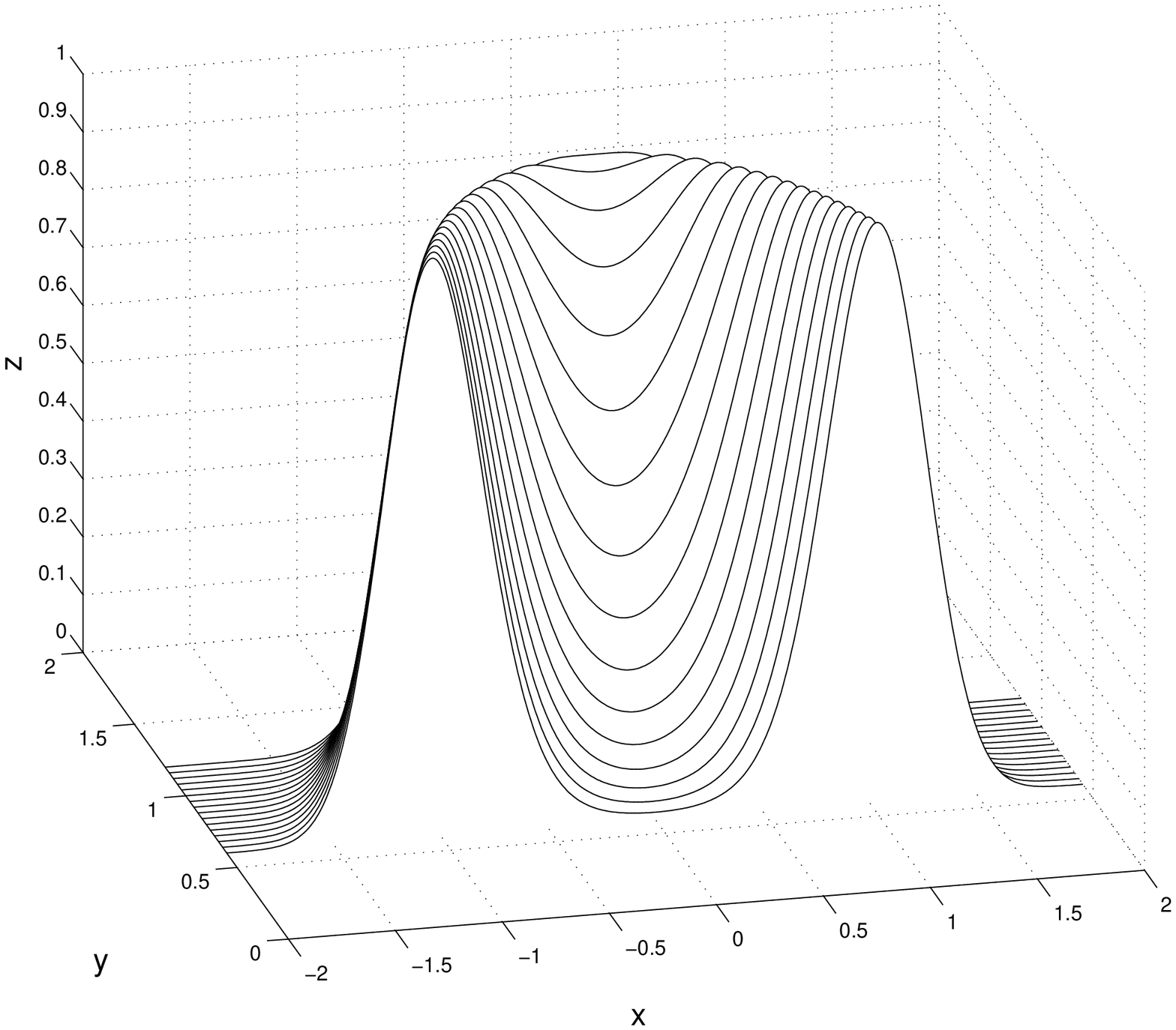}
\caption{Top: radiation conal component; bottom: cut-out section.
The height
shows the intensity of the radiation beam. The curves show sections along the
line of sight, i.e., pulse profiles. The bridge component is clearly seen.
\label{conal}}
\end{figure}

\begin{figure}[!htb]
\centering
\includegraphics[width=0.5\textwidth]{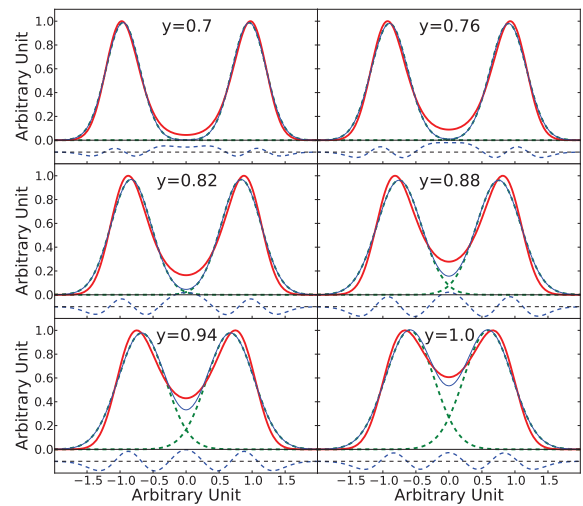}
\caption{Fits based on two Gaussian components of the sections in
  Fig.~\ref{conal} for different $y$ values. The thin lines (blue) are the sections. The dashed lines
(green) represent the two peaks, and the thick lines (red) trace the combination
of the two peaks. The blue dashed lines show the fit residuals, adopting the
same units as in the main panels.
\label{2peaks}}
\end{figure}

\begin{figure}[!htb]
\centering
\includegraphics[width=0.5\textwidth]{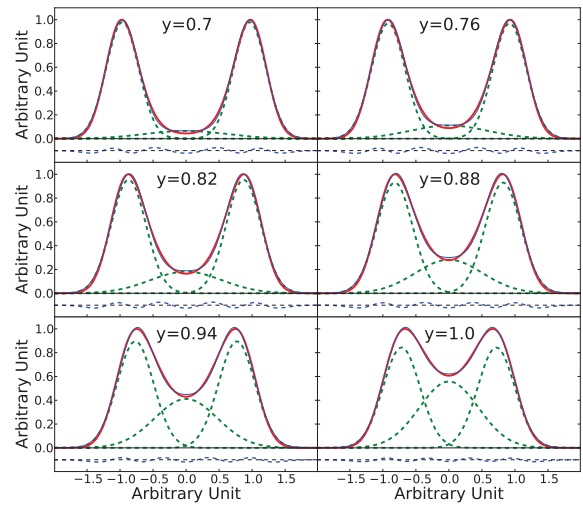}
\caption{As Fig.~\ref{2peaks}, but for fits with three
    Gaussian components.
\label{3peaks}}
\end{figure}

First, we will perform a numerical experiment pertaining to our
component analysis.
The intensity of a radiation beam composed of a single conal component
is related only to the angle between the radiation direction and the
magnetic axis (henceforth referred to as $\theta_{\mu}$), and its
radiation exists if and only if $\theta_{\mu}$ attains a value in a
specific narrow range.
A Gaussian intensity distribution, $F_{\mathrm{Gaussian}}$, is assumed
in the experiment,
$$
z=F_{\mathrm{Gaussian}}(\sqrt{x^2+y^2},A,r_0,\sigma),
$$ where
\begin{equation}
F_{\mathrm{Gaussian}}(x,A,x_0,\sigma)=
A\exp[-\frac{(x-x_0)^2}{2\sigma^2}].
\label{eq2}
\end{equation}
The model's 3D waterfall graph is shown in Fig.~\ref{conal}. The
parameters $A$, $r_0$, and $\sigma$ are set to 1.0, 1.2, and 0.2,
respectively.
The section planes of the conal component for different $y$ values are
similar to sections traversed by the radiation beam's line of sight
for different impact angles. The section profiles can be regarded as
pulse profiles.
From the section profiles, two peak components and a bridge component
between the two peaks can clearly be seen.
Upon trying to fit these profiles with both a two-peak model (see
Fig.~\ref{2peaks}) and a three-peak model (see Fig.~\ref{3peaks}), one
can see significant residuals and inconsistencies in the bridge region
for the two-component profile fit. (The Levenberg--Marquardt
  algorithm was used to fit the profiles, which is also adopted
  below.)
On the other hand, the three-peak model leaves small residuals, but
the width of the central peak is too large to treat it as an
independent component.
This characteristic may be an efficient tool to distinguish the bridge
component.

\subsection{Determining the Shape of the Conal component}

\begin{figure}[!tbh]
\centering
\includegraphics[width=0.5\textwidth]{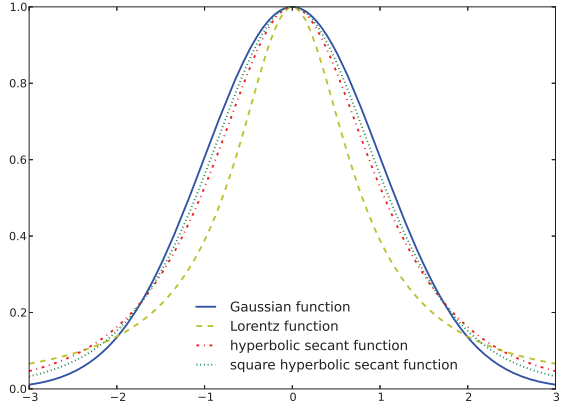}
\caption{Gaussian, Lorentz, hyperbolic secant, and square hyperbolic
  secant functions. In all cases, the location, height, and width are
  all set to 1.
\label{funcplot}}
\end{figure}

In the multi-component analysis of the pulse profile, a
Gaussian function with $\exp(-x^2)$ wing dampening is usually
used as the peak function~\citep{kram94,wu98}. However, the shape of the conal component in the real pulse
profile is as yet unknown. In this paper, to determine the most
probable shape of the conal component, the following four
functions are also used as trial peak functions to fit the pulse
profile:
\begin{mathletters}
\begin{equation}
F_{\mathrm{von~Mises}}(x,A,x_0,\sigma)=A\exp[\frac{1-\cos(x-x_0)}{\sigma^2}],
\label{eq3a}
\end{equation}
\begin{equation}
F_{\mathrm{sech}}(x,A,x_0,\sigma)=A\sech(\sqrt{\frac{\pi}{2}}\frac{x-x_0}{\sigma}),
\label{eq3b}
\end{equation}
\begin{equation}
F_{\mathrm{sech~square}}(x,A,x_0,\sigma)=A\sech^2(\sqrt{\frac{2}{\pi}}\frac{x-x_0}{\sigma}),
\label{eq3c}
\end{equation}
\begin{equation}
F_{\mathrm{Lorentz}}(x,A,x_0,\sigma)=\frac{A}{\frac{\pi}{2}(\frac{x-x_0}{\sigma})^2+1}.
\label{eq3d}
\end{equation}
\end{mathletters}
In Eq.~\ref{eq3a}, the factor has been chosen to make the function
resemble a Gaussian function, while in Eq.~\ref{eq3b}-\ref{eq3d}, the factors have been chosen such that all
functions have the same height, central position, and area as the
equivalent Gaussian function, since the same parameters are
used.
The von Mises function (Eq.~\ref{eq3a}) has also been used to fit the
pulse profile~\citep{manc12}, since it has a similar shape to the
Gaussian function and is periodic.
The hyperbolic secant function (Eq.~\ref{eq3b}) and the square
hyperbolic secant function (Eq.~\ref{eq3c}) naturally have exponential
wings.
These two functions also
show a peaked distribution; they have larger kurtosis and higher
wings than the Gaussian function.
The Lorentz function is characterized by inverse-square decreasing
wings.
The curves of Eq.~\ref{eq3b}-\ref{eq3d} and the
Gaussian function are shown in Fig.~\ref{funcplot}.
Since the von Mises function has a similar curve to the Gaussian
function, it has not been plotted.
As shown in Fig.~\ref{funcplot}, the Gaussian function has the
steeper wings than the square hyperbolic, the
hyperbolic, and the Lorentz functions.

\begin{figure}[!tbh]
\centering
\includegraphics[width=0.5\textwidth]{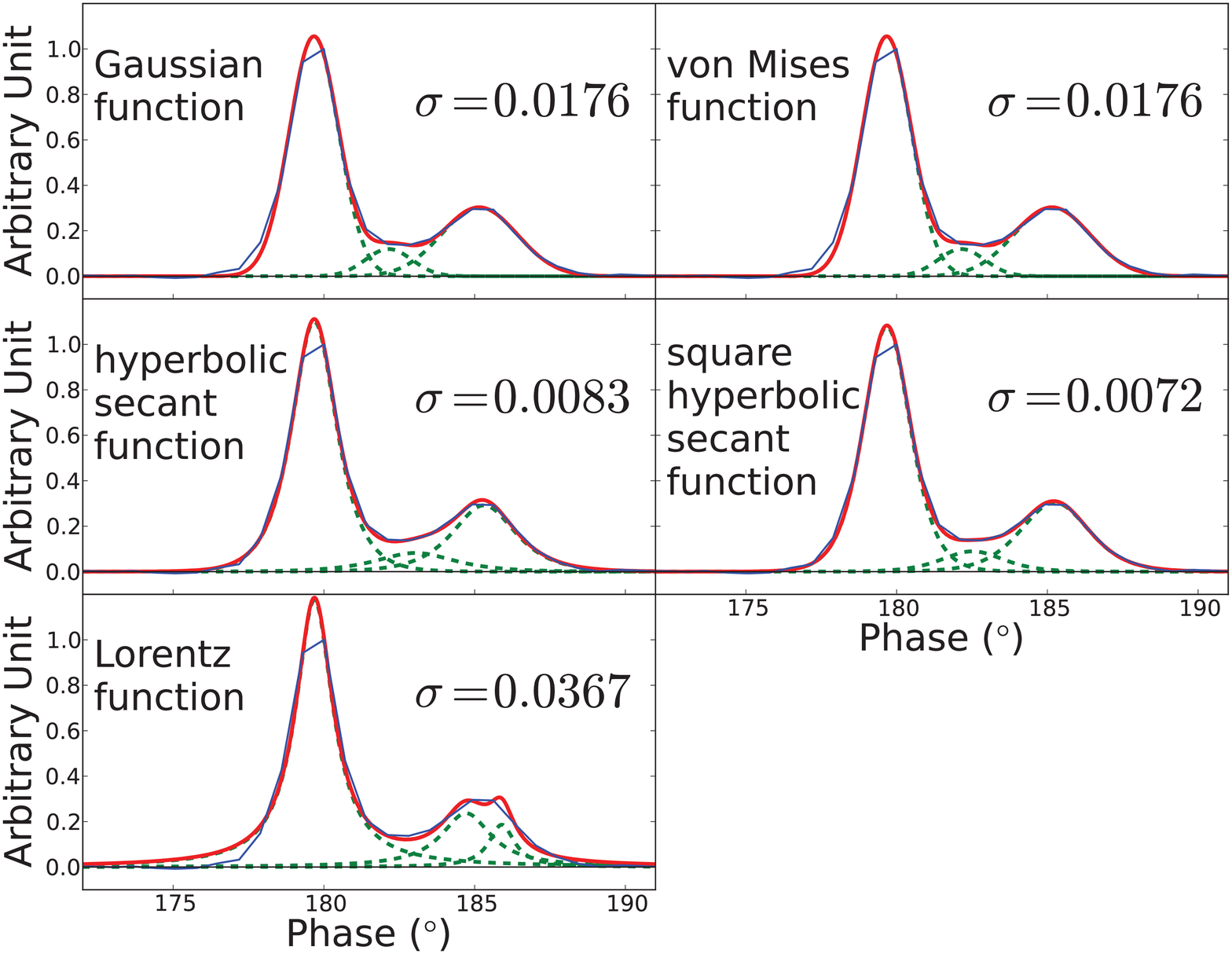}
\caption{Fits of the observed 2256 MHz profile of PSR B1133$+$16 with
  five different functions. The profiles have been aligned
    with the highest point of the first peak, and the pulse phase at
    the corresponding point is defined as 180$\degree$. The five
  panels separately show fits with a Gaussian function (left-hand
  panel, first row), a von Mises function (right-hand panel, first
  row), a hyperbolic secant function (left-hand panel, second row), a
  square hyperbolic secant function (right-hand panel, second row),
  and a Lorentz function (bottom panel). $\sigma$: root-mean-square of
  the residuals of the leading wings (for points with a pulse
    phase $< 179\degree.5$). The square hyperbolic secant
  function yields the best-fitting results in the leading-wing region,
  with the smallest $\sigma$.}\label{2256func}
\end{figure}

\begin{figure}[!tbh]
\centering
\includegraphics[width=0.5\textwidth]{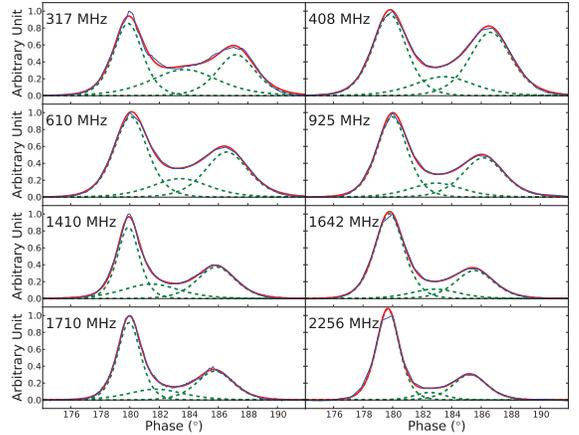}
\caption{Three-peak fits of the PSR B1133$+$16 multi-frequency
  profile. Only the profiles which include a bridge component have
  been taken into account in the fits. The green dashed lines
  highlight the three peaks, while the thick red line represents the
  combination of the three peaks. The thin blue lines are the observed
  profiles. The width of the central peak is too large to treat it as
  an independent component.
\label{1133fit}}
\end{figure}

\begin{table}[!tbh]
\begin{center}
\caption{Leading-wing residuals of the fits with five test functions.
  \label{bfit_resi}}
\begin{tabular}{cccccc}
\hline
\hline
Freq (MHz)  & G & M & H & S & L \\
\hline
317   &   0.0146   &   0.0146   &   0.0128   &   0.0096   &   0.0483   \\
408   &   0.0053   &   0.0053   &   0.0175   &   0.0064   &   0.0571   \\
610   &   0.0047   &   0.0047   &   0.0186   &   0.0063   &   0.0596   \\
925   &   0.0097   &   0.0097   &   0.0177   &   0.0083   &   0.0568   \\
1410   &   0.0121   &   0.0121   &   0.0081   &   0.0098   &   0.0341   \\
1642   &   0.0096   &   0.0096   &   0.0167   &   0.0064   &   0.0518   \\
1710   &   0.0099   &   0.0099   &   0.0085   &   0.0081   &   0.0299   \\
2256   &   0.0176   &   0.0176   &   0.0083   &   0.0072   &   0.0367   \\
Maximum   &   0.0176   &   0.0176   &   0.0186   &   0.0098   &   0.0596   \\
$\overline{\sigma^2}$ ($10^{-5}$)   &   12.6   &   12.6   &   20.2   &   6.2   &   230.7   \\
\hline
\end{tabular}
\end{center}
\footnotesize{Columns marked G, M, H, S, and L represent the
  leading-wing residuals of the fits with Gaussian, von Mises,
  hyperbolic secant, square hyperbolic secant, and Lorentz functions,
  respectively, and $\overline{\sigma^2}$ is the average value of the
  $\sigma^2$'s of all eight frequencies.}
\end{table}

As shown in Fig.~\ref{2256func}, these five functions have
  been used to fit the 2256 MHz pulse profile of PSR B1133$+$16.
Ignoring the trailing wings, which may be affected by interstellar
scattering, we have thus validated that exponential damping functions
such as the hyperbolic secant or square hyperbolic functions provide
better fits to the leading wings.
The leading-wing residuals of the multi-frequency profiles are
  shown in Table~\ref{bfit_resi}.
By comparing the root-mean-squares of the
leading-wing residuals, the fit based on the square
hyperbolic secant function has the smallest leading-wing
  residuals in most cases, and it also has the smallest maximum and
  average residuals.
Using the minimax criterion and Akaike's information criterion
  (AIC), which is defined as $\mathrm{AIC}=2k+\xi^2$, where $k$ is the
  number of parameters and $\xi^2$ is proportional to
  $\sigma^2$~\citep{burn02}, the square hyperbolic secant function is
adopted as the most appropriate shape of the conal component to fit
the pulse profile.
Because the square hyperbolic secant function is judged to be
the best peak function to fit the profiles, the PSR B1133$+$16
profiles are fitted with three square hyperbolic secant
peaks.
This fit yields a very wide central peak, similar to that experimented
(see Fig.~\ref{1133fit}; only the 8 frequency profiles that include a bridge
component have been taken into account in the fits), and it thus implies that the bridge
component of the PSR B1133$+$16 pulse profile cannot be independent.

\subsection{Profile Fits with a Conal Double Model}

\begin{figure}[tbh]
\centering \includegraphics[width=0.4\textwidth]{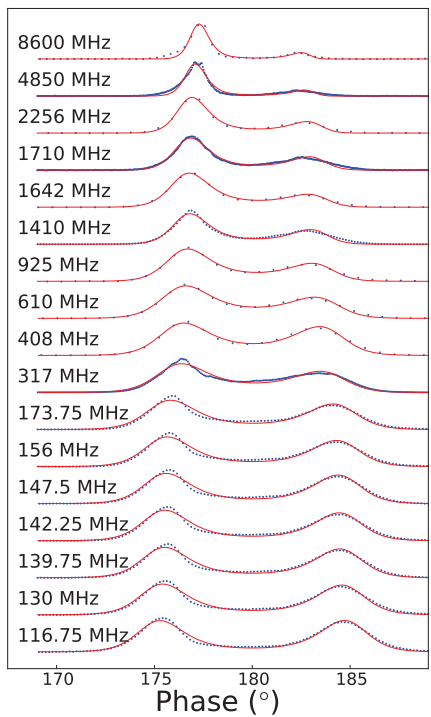}
\caption{Conal double model fits of the PSR B1133$+$16 multi-frequency
  profiles. The blue dotted lines are the observed profiles, while the
  red solid lines are the best-fitting results.}\label{bfit}
\end{figure}

\begin{figure}[tbh]
\centering
\includegraphics[width=0.4\textwidth]{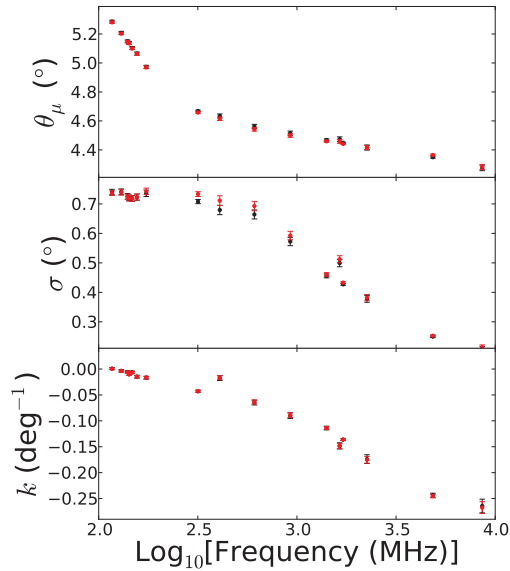}
\caption{Best-fitting parameters of the conal double model vs.
  frequency. From top to bottom, the parameters are the conal radius
  ($\theta_{\mu}$), the conal width ($\sigma$), and the conal
  asymmetry ($k$). The fit results of Eq.~\ref{eq1} are plotted as red
  points with error bars, and the black markers are the results of
  fits with a Gaussian-shaped conal component.
\label{para_vs_f}}
\end{figure}

In modeling the actual radiation beam, asymmetries caused by pulsar
rotation should be taken into account. This effect leads to
differences in the radiation intensity of the conal component
with pulse phase, $\phi$.
Here we only consider the first order (i.e., the linear variation) of
the radiation intensity.
In the conal double model, the conal component can be expressed as
\begin{equation}
I=F_{\mathrm{sech~square}}(\theta_{\mu},A,\theta_{\mu 0},
\sigma)\times(1+k\cdot(\phi-\phi_0)).
\label{eq1}
\end{equation}
Here, $\theta_{\mu}$ can be obtained from $\phi$~\citep{zhan07},
$$
\cos\theta_{\mu}=\cos\alpha\cos(\alpha+\beta)+
\sin\alpha\sin(\alpha+\beta)\cos(\phi-\phi_0),
$$
where $\alpha$ is the magnetic inclination angle and $\beta$ the
impact angle.
In Eq.~\ref{eq1},
$A$ is the amplitude of the Gaussian function,
$\theta_{\mu 0}$ and $\sigma$ are the angular radius and angular width
of the radiation conal component,
$k$ is the first order of the radiation intensity, which varies with
$\phi$ and also represents the conal component's asymmetry,
and $\phi_0$ is the phase measured when the line of sight, the
magnetic axis, and the rotation axis are coplanar.

\begin{table}[!tbh]
\centering
\caption{Fit parameters for the conal double model.
  \label{bfit_para}}
\begin{tabular}{cr@{$\pm$}lr@{$\pm$}lr@{$\pm$}l}
\hline
\hline
Freq (MHz)  & \multicolumn{2}{c}{$\theta_{\mu 0}$ ($^{\circ}$)} & \multicolumn{2}{c}{$\sigma$ ($^{\circ}$)} & \multicolumn{2}{c}{$k$ (deg$^{-1}$)} \\
\hline
116.75     &  5.283  &  0.008  &  0.737  &  0.008  &  0.001  &  0.002\\
130     &  5.205  &  0.008  &  0.739  &  0.009  &  $-$0.004  &  0.002\\
139.75     &  5.148  &  0.008  &  0.724  &  0.009  &  $-$0.006  &  0.002\\
142.25     &  5.137  &  0.008  &  0.719  &  0.009  &  $-$0.010  &  0.002\\
147.5     &  5.102  &  0.008  &  0.718  &  0.009  &  $-$0.007  &  0.002\\
156     &  5.064  &  0.009  &  0.72  &  0.01  &  $-$0.015  &  0.002\\
173.75     &  4.971  &  0.009  &  0.74  &  0.01  &  $-$0.017  &  0.002\\
317     &  4.662  &  0.006  &  0.723  &  0.008  &  $-$0.043  &  0.002\\
408     &  4.62  &  0.01  &  0.70  &  0.02  &  $-$0.017  &  0.004\\
610     &  4.55  &  0.01  &  0.68  &  0.02  &  $-$0.064  &  0.004\\
925     &  4.51  &  0.01  &  0.58  &  0.01  &  $-$0.090  &  0.005\\
1410     &  4.463  &  0.007  &  0.459  &  0.006  &  $-$0.114  &  0.003\\
1642     &  4.47  &  0.01  &  0.51  &  0.01  &  $-$0.148  &  0.005\\
1710     &  4.444  &  0.003  &  0.431  &  0.003  &  $-$0.136  &  0.002\\
2256     &  4.42  &  0.02  &  0.38  &  0.01  &  $-$0.174  &  0.007\\
4850     &  4.358  &  0.006  &  0.251  &  0.003  &  $-$0.245  &  0.003\\
8600     &  4.28  &  0.02  &  0.211  &  0.008  &  $-$0.27  &  0.01\\
\hline
\end{tabular}
\end{table}

In the fit, we adopt inclination and impact angles of $51\degree.3$
and $3\degree.7$, respectively~\citep{lyne88}.
The resulting pulse profile fit for PSR B1133$+$16 with this conal
double model is shown in Fig.~\ref{bfit}.
The bridge components could clearly be fitted adequately.
The residuals in excess of the uncertainties in the data are caused by
both our adoption of a square hyperbolic secant profile and
our assumption of linear variation of the radiation
intensity. Aberration and retardation may also affect the shape of the
conal component~\citep{kuma13}.
The error due to model selection bias is much larger than the
  system noise, which shows that the system noise does not affect our
  analysis.
The fit parameters as a function of frequency are included in
Table~\ref{bfit_para} and shown in Fig.~\ref{para_vs_f}.
The uncertainties in the parameters were calculated based on
application of the Jacobian matrix around the best-fitting solution.
Since the amplitudes ($A$) and initial positions ($\phi_0$) of the
profile are not important, we show only the other three parameters.
In Fig.~\ref{para_vs_f}, the parameters obtained by performing fits
with a Gaussian-shaped conal component are also shown for comparison.
They are approximately the same as for the fit obtained with a
square hyperbolic secant profile.
As shown in this figure, the opening angle of the radiation
  cone decreases with frequency, which means that the phase difference
  between two peaks increases.
This phenomenon occurs for most pulsars~\citep{chen14}.
In following, only the results calculated based on the square
hyperbolic secant profile are adopted.

\section{Relativistic Particles in the Magnetosphere}

\subsection{Radiation altitude}

\begin{figure}[!tbh]
\centering
\includegraphics[width=0.4\textwidth]{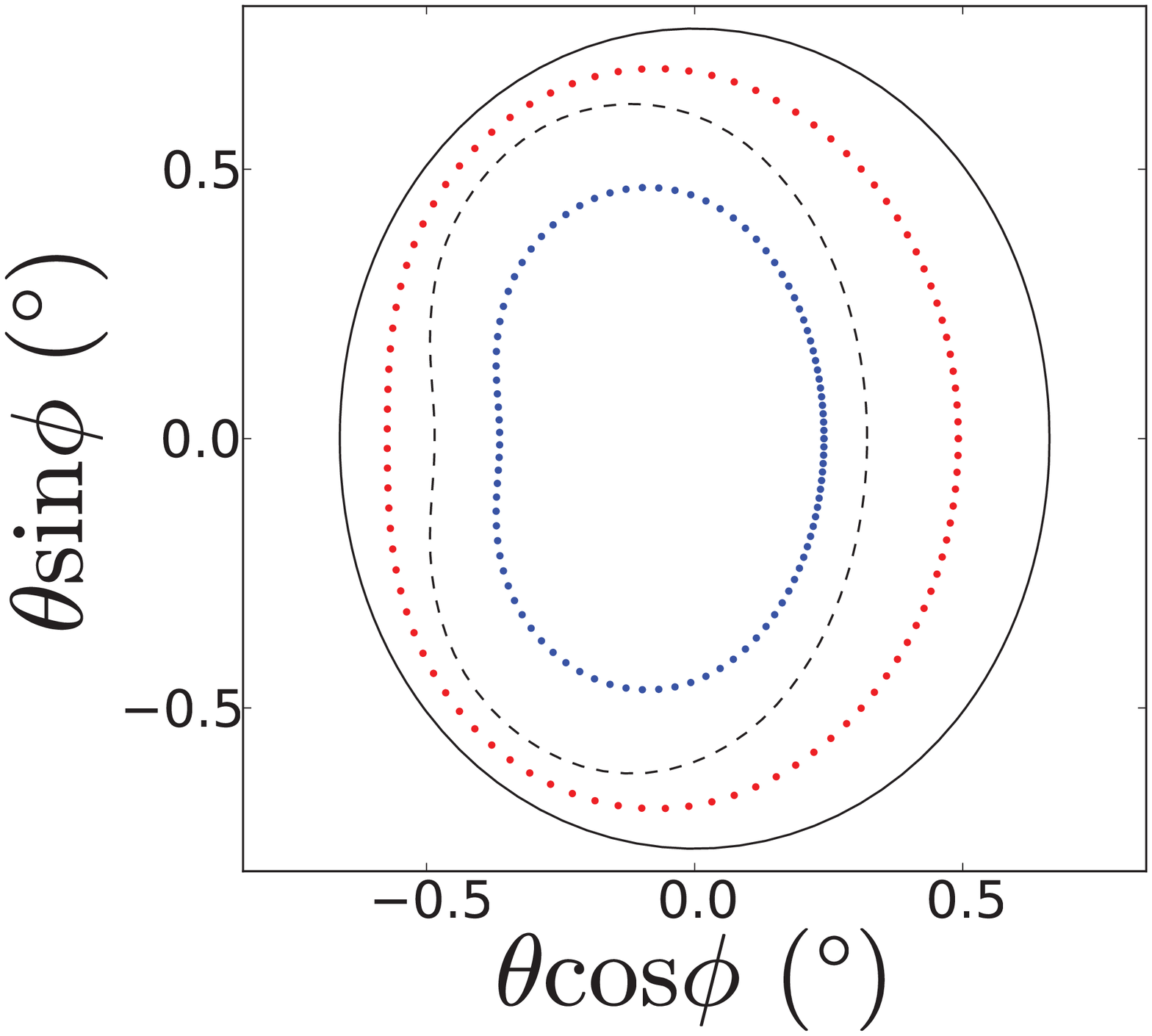}
\includegraphics[width=0.4\textwidth]{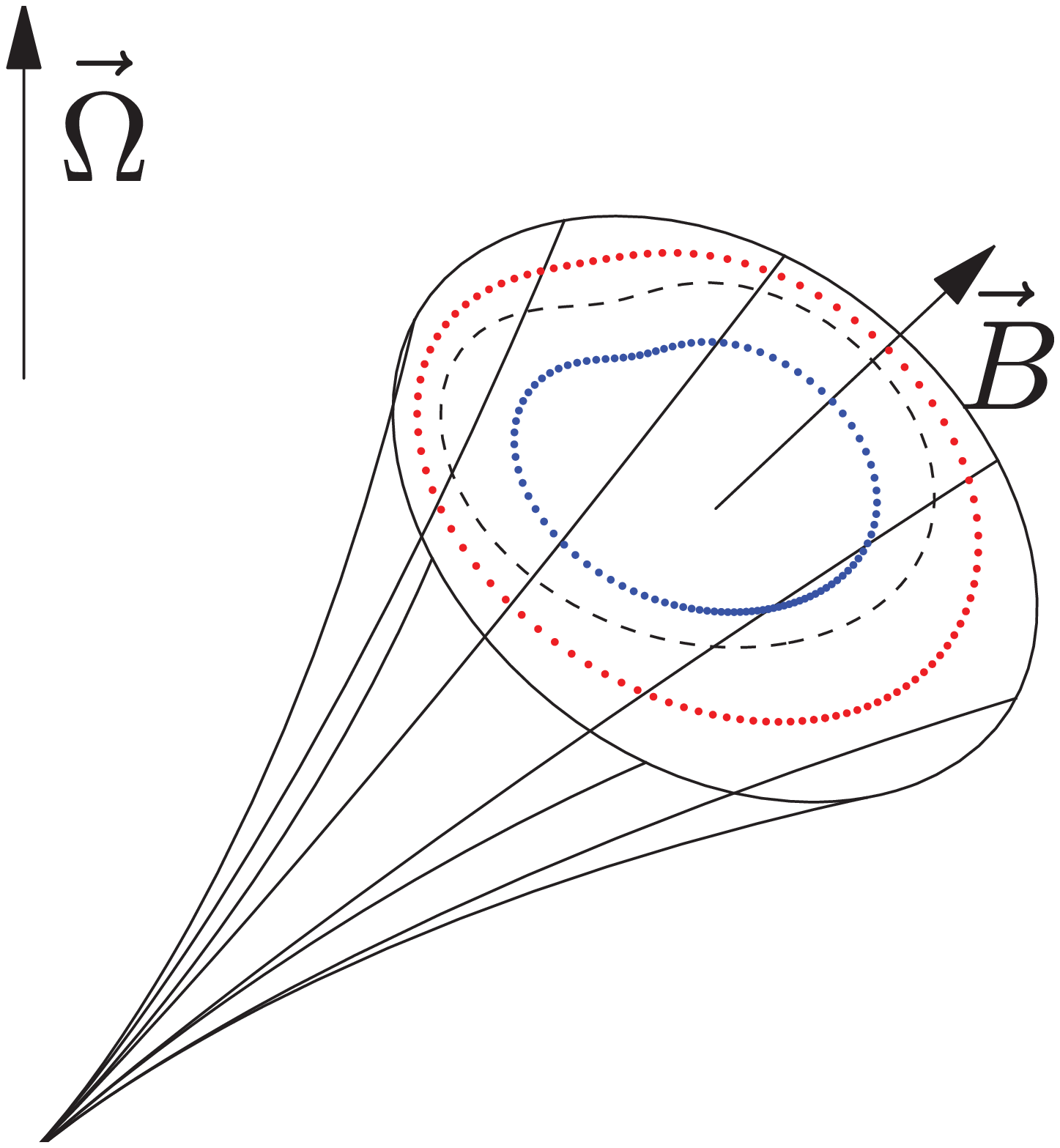}
\caption{Field line in the core region (blue dotted line) and in the annular
region (red dotted line). The black solid curve is the last opening field line, and
the black dashed line is the critical field line. Top: section in the polar region;
bottom: 2D radiation beam.
\label{icgiag}}
\end{figure}

Before analyzing the distribution of the particle energies in the
magnetosphere, the radiation altitude for each frequency band must be
calculated.
the radiation altitude can be calculated under the assumption of a dipole field and also assuming that the
radiation is emitted near a specific field line, since the pulse phase
$\phi$ is fixed.
First, the equation describing the dipole field line can be written as
\begin{equation}
r=R_{\rm e}\sin^2\theta,
\label{eq4}
\end{equation}
where $r$ and $\theta$ are spherical coordinates with the magnetic
axis in the zenith direction and $R_{\rm e}$ is the field line's
diameter. The latter is a fixed parameter for a specified field line.
This specified field line could be either in the core region or in an
annular region~\citep{qiao04a,qiao07}.
The field line in the core region is set at 0.75 times the critical field
line (which passes the point of intersection between the null charge
surface and the light cylinder) for the zenith angle of the point of
intersection between the field line and the pulsar surface.
The field line in the annular region is set at the average value
of the critical field line and the last opening field line (which is
tangential to the light cylinder, and which can be calculated
following the method of \cite{zhan07}).
All of these field lines are shown in Fig.~\ref{icgiag}.
The dotted line shows the real shape of the radiation beam (the core
region's field line is marked blue, while the annular region's field
line is marked red).
The last opening field line is shown as the black solid line and the
critical field line is rendered as the black dashed line.
In the top panel, a section of the field lines in the polar-cap region
is shown, while a 2D sketch of the radiation beam is offered in the
bottom panel.
The detailed calculation of $R_{\mathrm{e}}$ of the critical field
line is given in Appendix A.

\begin{figure}[!tbh]
\centering
\includegraphics[width=0.4\textwidth]{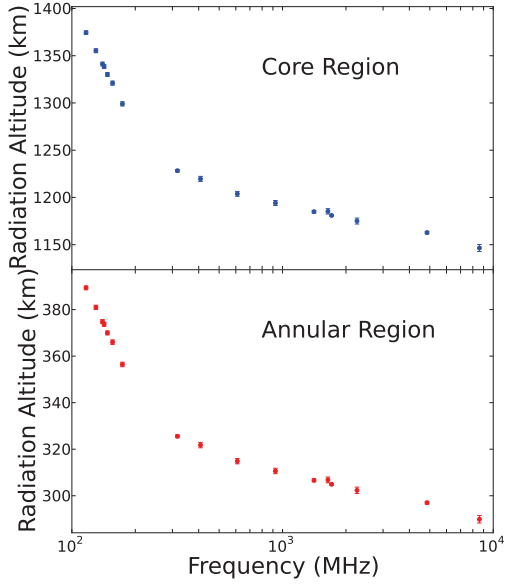}
\includegraphics[width=0.4\textwidth]{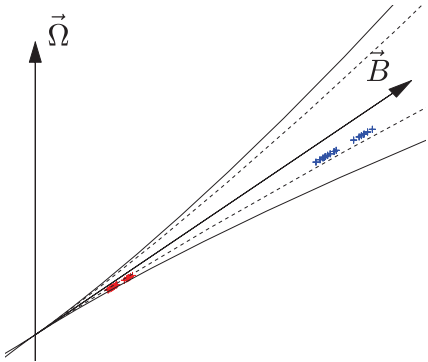}
\caption{Radiation altitude. The blue markers show the results for the core
region, while the red markers show the results pertaining to the annular region.
The plane defined by the rotation and magnetic axes is shown in the bottom
panel. The black solid curve represents the last opening field line, while the
black dashed line shows the critical field line.
\label{rad_posi}}
\end{figure}

Combined with the assumption that the radiation is emitted along the
tangent line of the magnetic field, the relation between $\theta$ and
$\theta_{\mu}$ can be obtained,
\begin{equation}
\tan\theta_{\mu}=\frac{3\sin 2\theta}{1+3\cos 2\theta}.
\label{eq5}
\end{equation}
With Eq.~\ref{eq4} and Eq.~\ref{eq5}, the radiation's position can be
determined: see Fig.~\ref{rad_posi}.
In the top panel, the radiation altitude for different radiation
regions is shown, revealing that higher-frequency photons come from
lower positions in the pulsar.
In the bottom panel, the relative radiation altitudes for different
radiation regions are plotted.
It is apparent that the radiation position in the core region's model
is significantly higher than that in the annular region's model.

\subsection{Particle Energy Loss}

\begin{figure}[!tbh]
\centering
\includegraphics[width=0.5\textwidth]{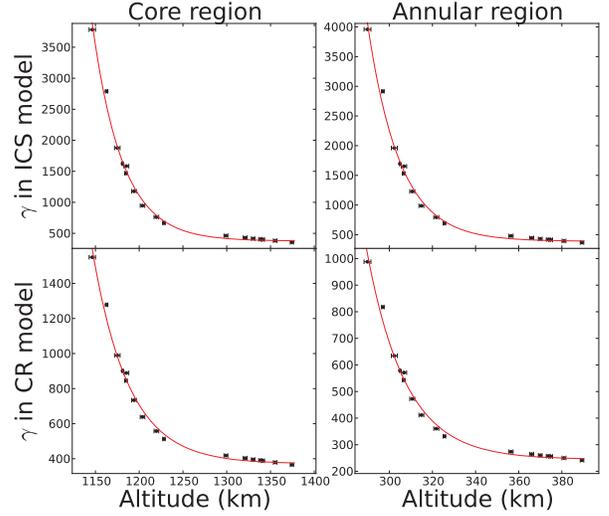}
\caption{Evolution of particle energy with altitude for different
  cases. Fits to the data points using Eq.~\ref{eq8} are
  also shown.
\label{lorentz_factor}}
\end{figure}

In both the ICS and CR models, the energy of the radiating
  particles can be calculated if the radiation frequency and position
are known.
The radiation frequency in the CR model, $\nu_{\mathrm{CR}}$, can be
calculated from the Lorentz factor of the particles, $\gamma$, and the
curvature radius, $\rho$, of the field line \citep{rude75},
\begin{equation}
\nu_{\mathrm{CR}}=\frac{3}{2}\gamma^3\frac{c}{2\pi\rho},
\label{eq6}
\end{equation}
where $c$ is the speed of light and
$$
\rho\approx\frac{4r}{3\theta},
$$
and this approximation applies to the case
$\theta\ll\frac{\pi}{2}$.
Based on Table.~\ref{bfit_para} and Eq.~\ref{eq5}, $\theta$ could
be still smaller than $4^{\circ}$, and the approximation above should be
always valid.
The particle energy can then be obtained in the CR model.
In the ICS model, the radiation frequency, $\nu_{\mathrm{ICS}}$, can
be expressed~\citep{zhan07} as
\begin{equation}
\nu_{\mathrm{ICS}}=2\nu_0\gamma^2(1-\eta\frac{M}{N}),
\label{eq7}
\end{equation}
where $\nu_0$ is the initial frequency of the photons
  generated by gap sparking, which is about 1 MHz before
  scattering, $\eta=\frac{v}{c}$, and
$$
M=2r\cos\theta-R[3\cos\theta\sin\theta\sin\theta_c\cos(\psi-\psi_c)+
$$
$$
(3\cos^2\theta-1)\cos\theta_c],
$$
$$
N=(1+3\cos^2\theta)^{1/2}\{r^2+R^2-2rR[\cos\theta\cos\theta_c+
$$
$$
\sin\theta\sin\theta_c\cos(\psi-\psi_c]\}^{1/2},
$$
where $\theta_c$ and $\psi_c$ are the polar coordinates of the initial
photon-radiation position and $R$ is the radius of the pulsar.
The particle energy can then also be obtained in the ICS model.
The results are shown in Fig.~\ref{lorentz_factor}, which suggests
that the particle energy should decrease with altitude in both the ICS
and the CR models, i.e., the particles would lose energy while moving
along the magnetic field lines.
This result is consistent with the assumptions adopted in some previous
studies regarding the ICS model~\citep{qiao98, qiao01, zhan07}.

\begin{table}[!tb]
\centering
\caption{Results of fits to the data points in
  Fig.~\ref{lorentz_factor} using Eq.~\ref{eq8}.
  \label{gammafit_para}}
\begin{tabular}{c|r@{$\pm$}lr@{$\pm$}lr@{$\pm$}l}
\hline
\hline
Case  & \multicolumn{2}{c}{$\gamma_1$} & \multicolumn{2}{c}{$A$} & \multicolumn{2}{c}{$H_0$ (km)}  \\
\hline
core (ICS) &  380  &  70  &  1.18  &  0.07$\times10^{18}$  &  34.28  &  0.01  \\
core (CR)  &  370  &  30  &  1.24  &  0.07$\times10^{15}$ &  41.46  &  0.02  \\
annular (ICS)  &  390  &  80  &  1.01 &  0.06$\times10^{12}$  &  14.92  &  0.02  \\
annular (CR)  &  240  &  20  &  7.8  &  0.5$\times10^9$  &  17.97  &  0.03  \\
\hline
\end{tabular}
\end{table}

As shown in Fig.\ref{lorentz_factor}, the rate of decrease of the
particle energy also decreases with altitude.
The functional form describing the particle energy as a function of
altitude $H$ is suggested to be exponential, as was also adopted in a
number of previous studies~\citep{wang13,zhan07},
\begin{equation}
\gamma=A\exp(H/H_0)+\gamma_1.
\label{eq8}
\end{equation}
The points in Fig.~\ref{lorentz_factor} are fitted with this function;
the result is shown in Table~\ref{gammafit_para} and it is also
presented in the figure.
Note that $A$ is not the initial lorentz factor of particles
but just a fitting parameter, since the Lorentz factor of particles
beyond the radiation region that generates emission with the frequencies
chosen above is hard to estimate.
From the figure, the Lorentz factors of the particles have a lower
limit of $\sim$300. In other words, only high-energy particles would
lose energy in the magnetosphere.
This phenomenon cannot be explained by other, traditional
magnetosphere models.

\subsection{Particle Energy Dispersion}

\begin{figure}[!tbh]
\centering
\includegraphics[width=0.5\textwidth]{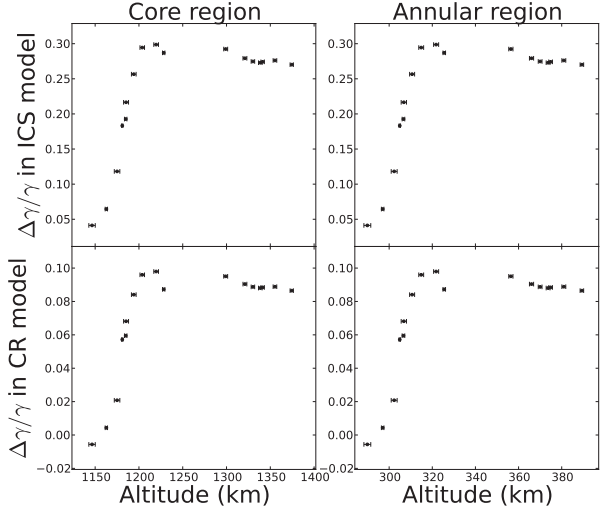}
\caption{Particle energy dispersion vs. altitude, represented by
  $\Delta\gamma/\gamma$.
As shown in Fig.~\ref{lorentz_factor} and here, the variation in
particle energy is similar to that relating to the particle-energy
dispersion.
The particle energy decreases rapidly while the particle energy
dispersion increases sharply. On the other hand, as the rate of
decrease becomes smaller, the dispersive process comes to a halt.
These particle-energy properties provide constraints on the physical
processes in the magnetosphere.
\label{dgamma}}
\end{figure}

In the CR model, the conal beam width can be derived based on
Eq.~\ref{eq4}-\ref{eq6}, and the assumption
$\theta\ll1$, i.e.,
\begin{equation}
\frac{\Delta\theta_{\mu}}{\theta_{\mu}}\approx\frac{\Delta\nu}{\nu}+3\frac{\Delta\gamma}{\gamma},
\label{eq9}
\end{equation}
where $\Delta\theta_{\mu}=2.012\sigma$ is the full width at half
maximum of the conal component, $\Delta\nu$ is the bandwidth of each
band, and $\Delta\gamma$ is the dispersion of $\gamma$.
Note that $\Delta\gamma$ is not the Lorentz factor dispersion of the
particles at a certain location, but the Lorentz factor dispersion of
the particles that emit photons with the same phase.
From Eq.~\ref{eq9} it follows that, for a specified frequency, the
conal width can be determined from the radiation field line and the
particle energy dispersion.
As the field line does not change with radiation frequency,
any change
in the conal width as the frequency increases is due to variation
in the particle energy dispersion with altitude.

In the ICS model, the conal width can also be derived with
Eq.~\ref{eq7} and the assumptions $\theta\ll1$, $\theta_c\ll1$, and
$R\ll r$,
\begin{equation}
\frac{\Delta\theta_{\mu}}{\theta_{\mu}}\approx\frac{1}{2}\frac{\Delta\nu}{\nu}+\frac{\Delta\gamma}{\gamma}.
\label{eq10}
\end{equation}
Similarly, the conal width changes as a result of the variation in
particle energy dispersion.
With Eq.~\ref{eq9} and Eq.~\ref{eq10}, $\Delta\gamma$ can be
calculated. The results are shown in Fig.~\ref{dgamma}.
In this figure and below, the particle energy dispersion is expressed
as $\frac{\Delta\gamma}{\gamma}$.

In fact, radiation emitted in same phase of the profile also originates
from different field lines. This means that $\Delta\gamma$ does not
only include the width of the particle spectrum, but also contains
energy variations around the locations calculated in Section 4.1.
In this case, the latitudinal distribution of the particles would also
contribute to $\Delta\gamma$.
Under the assumption of the presence of a core gap or an annular gap,
the initial latitudinal distribution of particles and particle energy
should be consistent. In other words, the latitudinal difference of
particles in the vicinity of the radiation location is equivalent to
the latitudinal energy dispersion.

As shown in Fig.~\ref{dgamma}, the particle energy disperses as
altitude increases in the lower region, but in higher regions it does
not yet change.
This change in energy dispersion is similar to that of the energy-loss
process, which is analyzed in Section 4.2. This may imply that the
decrease and dispersion in
particle energy are caused by the same physical process.
From Fig.~\ref{dgamma} it follows that, in the ICS model, the particle
energy dispersion could still be positive, and the negative particle
dispersion at lower altitudes cannot be explained if the CR model is
adopted.
The particle energy dispersion could be used to determine that there
is weak bridge component at both high and low frequencies.
For low frequencies, the conal radius is located far above the impact
angle, and measurements along the line of sight would produce two
almost independent peaks.
However, at high frequencies, the conal component is narrower because
of the narrower particle energy spectrum, and the bridge component
would also disappear.
This is also the reason why only the profiles observed in
mid-frequencies have obvious bridge components, and thus 
why eight profiles are chosen in Fig.~\ref{1133fit}.

\section{Discussion}

The pulse profile of PSR B1133$+$16 has two obvious peaks.
Assuming that these two peaks are two independent components,
the profile should be fitted by adopting a two-peak
model. In fact, the bridge region between the two peaks cannot
be adequately fitted in the two-peak model, which implies that
the two peaks cannot be regarded as independent components.
Three-peak model fits yield a central component that is too
wide in the bridge region, which means that the bridge
component could be an artificial one. On the other hand, an
emission model with more than three components is too
complex for this source and cannot be made to agree with core
or conal emission geometries. In addition, the profile can be
fitted very well with a 2D conal function, which has fewer
parameters (five) than even the two-peak model (six parameters).
It also implies that the two peaks should be the
natural result of real conal components of radiation.

In Section 4, we demonstrated that radiation of different
pulse phases is related to the particle energy at the radiation
position and that the widths of peaks can be understood as the
width of the particle energy spectrum, to some extent. If so, the
flux of the wings versus the pulse phase could be naturally
regarded as the manifestation of a particle energy spectrum. In
the ICS and CR models, the Gaussian peak may imply that
particles at the radiation position have an energy spectrum
resembling
$\exp(-\gamma^2)$, but there is no reasonable physical
process that could result in such a spectrum. Conversely, a
power-law or exponential spectrum could be understood more
easily. Peak functions with power-law or exponential wings,
such as a Lorentz function, a hyperbolic secant function, or a
square hyperbolic secant function, may be better physically
motivated to fit the pulse profile, and fits using a square
hyperbolic function indeed yield a better result. The angular
distribution of the particles could also affect the damping rate
of the wings, and whether the particle spectrum or its angular
distribution plays a leading role in this aspect needs to be
studied in more detail.

The notion that peak functions with exponential wings lead
to better fit results, as mentioned in Section 3, also applies to
other pulsar profiles. In other words, the pulse profile
components may have a shape similar to the square hyperbolic
secant function. This insight should be considered in pulsar
timing. For example, while constructing a profile template, the
square hyperbolic secant function could be used to fit the
template profile instead of Gaussian or von Mises functions.

As shown in Fig.~\ref{rad_posi}, the range of radiation
  altitudes for different frequencies depends significantly on the
  assumption adopted for the radiation region.
The difference in radiation altitude between 116.75 and 8600 MHz is
228 km for the core-region radiation assumption and only 100 km if we
assume that the radiation originates in an annular region.
As a consequence, a time lag of 0.76 ms is predicted between 116.75
and 8600 MHz for core-region radiation, but only 0.33 ms for
annular-region radiation.
Such effects would significantly affect simultaneous
multi-frequency pulsar timing results if the same template is used
for profiles at different frequencies in pulsar timing. In other
words, timing may be an effective method to distinguish the
location of the radiation region, but the dispersion measure (DM)
should first be determined accurately.

The positive correlation between the absolute value of the conal
asymmetry and the pulsar frequency is shown in Fig.~\ref{para_vs_f}.
It is apparent that the asymmetry varies continuously as the
radiation localization increases in altitude. It leads to a smaller
and smaller trailing peak (relative to the leading peak) as
frequency increases, and in a very high frequency band the
trailing peak could be predicted to be vanishing. Such
asymmetry may be caused by the inhomogeneous distribution
of the parallel electric field in the polar gap, which may be a
result of pulsar rotation, and the density and spectrum of the
particles would be different for different azimuth angles of the
conal component. In addition, in the context of the ICS model,
the conal asymmetry could also be due to the asymmetry of the
initial photon energy distribution. If the luminosity of the initial
photons in the polar-cap region is inhomogeneous, the radiation
flux density would vary for different magnetic field lines. The
inhomogeneity of the photon field weakens with increasing
altitude, because the differences in the distance between the polar cap and different radiation locations become smaller. The
asymmetry of the pulse profile would thus disappear. While
this effect is reflected in the pulse profiles, the asymmetry
would increase with increasing frequency, which is consistent
with the observations.

Other effects, such as field-line bending caused by pulsar
  rotation, aberration, retardation, and plasma effects, could also
  affect the shape of the profile.
The first three effects would affect the arrival time of the
radiation, and the time difference $\Delta t$ characteristic of all
three effects can be estimated: $\Delta t\sim\Delta h/c$, where
$\Delta h$ is the altitude difference between the radiative location in a
given band.
This altitude difference could also be estimated as $\Delta h\sim
r\frac{\Delta\theta_{\mu}}{\theta_{\mu}}\sim
r\frac{\Delta\gamma}{\gamma}$.
If the radiation is generated in the core region, $\Delta t\sim
1~\mathrm{ms}$, while for an annular region, $\Delta t\sim
0.1~\mathrm{ms}$.
Given that the period of PSR B1133$+$16 is 1.1879~s, these three effects
cannot significantly affect the observed profile.
The uneven plasma distribution in the magnetosphere could lead to
light deflection, which would result in deformation of the pulse
profile.
Using the laws of reflection, the deflection angle $\Delta\theta$
could be derived, $\Delta\theta\sim\theta_{\mu}\frac{\nu_{\rm
    pe}^2}{\nu\nu_{\rm ce}}\frac{\Delta\gamma}{\gamma}$, where
$\nu_{\rm pe}$ is the electron plasma frequency and $\nu_{\rm ce}$ is
the electron cyclotron frequency at the radiation location.
Adopting the Goldreich--Julian density~\citep{gold69} as the density
of the free charged particles in a plasma, this deflection angle can
be calculated.
For the core, the deflection angle can be up to $\sim4.3\times10^{-9}$
if the lowest frequency 116.75 MHz is adopted, while for the annular
region, this value is $\sim1.4\times10^{-8}$.
Even if the magnetic field is ignored, the deflection angle which is
expressed as $\Delta\theta\sim\theta_{\mu}\frac{\nu_{\rm
    pe}^2}{\nu^2}\frac{\Delta\gamma}{\gamma}$ attains values of
$5.7\times10^{-5}$ and $9.4\times10^{-4}$ for the two cases.
These results show that plasma effects are insignificant for our
analysis of the profiles of PSR B1133$+$16.
As discussed above, these effects would not change the distance
between the two peaks or their widths significantly, so that they will
not affect the results obtained in the Section 4.

The initial frequency of the photons before scattering
$\nu_0\sim1~\mathrm{MHz}$ is decided by both theories and observations.
From theories in the RS model~\citep{rude75}, a low
frequency wave is produced in the inner gap sparking, which means the
timescale should be $\sim10~\mathrm{\mu s}$.
In the RS model, only the curvature radiation
is taking into account, but near the star surface the strong magnetic
field effects should be taken into account.
When such effects are considered, the gaps should be
nearly 5 times lower than RS's~\citep{zhan97}.
From observations, the occasional sub-pulses with a time structure
shorter than $10\mathrm{\mu s}$~\citep{hank71,bart82,hank03} also shows that the
inner gap should be lower than RS's.
With this evidence, the initial frequency is set to be
$\nu_0=10^6~\mathrm{Hz}$, which is $2\pi$ times larger than the
choice of~\cite{zhan07}.
This choice means that the sparking timescale is shorter than
$10~\mathrm{\mu s}$ and it fits both theories and observations better.

$\gamma$-ray radiation from pulsars is suggested to originate from the
annular and core regions~\citep{du10,du11}, while radio radiation
coming from the annular region has also been previously
discussed~\citep{qiao07}.
For PSR B1133$+$16, without considering additional
acceleration effects, the particle energy in the acceleration region
can be obtained by tracing back the energy-loss rate to the pulsar
surface. The particle energy will be too large if the emission comes
from the core region.
This may imply that the radiation comes from the annular region.
Theoretically, high-energy particles are preferably generated in the
inner annular gap for short-period pulsars, since the annular gap
region is so large that it has a large electric potential drop to
accelerate charged particles~\citep{du10}.
However, the ``bi-drifting'' phenomenon of PSR J0815$+$0939, which has
a period $\sim 0.6452~\mathrm{s}$~\citep{cham05}, may indicate that
radio emission from a pulsar with a slightly longer period could also
be generated in the annular region~\citep{qiao04b}.
As a pulsar with a period of $\sim 1.1879~\mathrm{s}$, PSR B1133$+$16
could be a long-period pulsar candidate that emits radio radiation
from its annular region.
Nevertheless, considering the acceleration processes in the
magnetosphere~\citep{hard08}, the conclusion obtained above should be
treated with caution.

The double-peaked profile could also be understood in the context of
the fan-beam model~\citep{wang14}, and the two
peaks could be considered to come from two different, patchy
beams. However, the wide-band, multi-frequency properties of
the pulse profile in the fan-beam model have not been fully
clarified, so that further theoretical work is required.

The luminosity and spectrum are also important to
distinguish among the different emission mechanisms. A
simple simulated pulse profile in both the ICS and CR models
is considered in Appendix B. In the simulation, we assume
completely coherent radiation. However, in a real-world
situation, the degree of coherence of the radiation should not
be 100\%. It is a time-dependent function of both the electron
and radio-flux densities, which is difficult to calculate. This
causes significant difficulties for simulations of pulsar spectra.
Such a study may be done in a future paper. In fact, phaseresolved
spectra of PSR B1133$+$16 have been studied~\citep{chen07}. These would provide powerful observational
evidence to test different models for the radiation mechanism.

\section{Conclusions}

By fitting the profile of PSR B1133$+$16 with three
independent components, we found that the central peak is
too wide to be regarded as an independent component.
Together with a numerical experiment, the radiation beam of
PSR B1133$+$16 is determined to be of conal type.

A 2D beam function instead of a simple Gaussian function
was used to fit the multi-frequency profile of PSR B1133$+$16,
and correlations between the radiation beam parameters and the
observed frequency were obtained. The radiation altitudes of
multi-frequency bands were also calculated. With the radiation
frequency and the corresponding altitude, the particle energy has
been derived based on two radiation models. In the ICS and CR
models, the particle energy is found to decrease rapidly while it
is high, but this decreasing tendency flattens as the particle
energy drops to $\sim$300~MeV.
The particle energy dispersion is
also derived, and the dispersivity has a negative relation with the
particle energy. When particles flow out along the field line
(losing energy), both the ICS and CR models could reproduce
the curve of the relation between the opening angle of the
radiation cone ($\theta_{\mu 0}$) and the frequency for PSR B1133$+$16. More
observations with high time resolution are needed to distinguish
between these two radio radiation models.

Peak functions with exponential wings could fit our profiles
better, and this approach could be useful for constructing a
template profile for pulsar timing. It could also imply an
exponential particle spectrum to some extent, but data with
higher time resolution are needed to test this conjecture.

\acknowledgments
This work was supported by the National Basic Research
Program of China (973 program, grant No. 2012CB821800),
the talent team of science and technology innovation in
Guizhou Province (Grant No.(2015)4015) and
the National Natural Science Foundation of China (grant Nos
11373011, 11303069, 11225314, 11103021, 11565010 and 11165006).
We would like to thank Dr. Dipanjan Mitra for sharing the 317 MHz pulse profile data
observed with the GMRT and Dr. Ramesh Karuppusamy for sharing the 116.75--173.75 MHz
pulse profile data observed with the Westerbork Synthesis Radio Telescope (WSRT).
We would also like to thank the Kunming 40~m radio telescope,
and EPN teams for providing their pulse profiles.
In particular, we would like to greatly thank Prof. Richard de Grijs
for his kind help on improving the language carefully.
We sincerely thank the anonymous referee for insightful
suggestions and valuable comments.

\appendix

\section{A.~~Calculating the critical field line}

In magnetic coordinates $(r,\theta,\phi)$, which treat the magnetic
axis as the direction of the zenith, the magnetic dipole field can be
presented as follows,
\begin{mathletters}
\begin{equation}
B_r=\frac{B_0}{r^3}\cos\theta,
\label{br0}
\end{equation}
\begin{equation}
B_\theta=\frac{B_0}{2r^3}\sin\theta,
\label{bt0}
\end{equation}
\begin{equation}
B_\phi=0.
\label{bp0}
\end{equation}
\end{mathletters}
Transforming these into rotational coordinates $(r',\theta',\phi')$,
whose zenith is the rotation axis,
\begin{mathletters}
\begin{equation}
B_{r'}=\frac{B_0}{r'^3}(\cos\theta'\cos\alpha+\cos\phi'\sin\theta'\sin\alpha),
\label{br1}
\end{equation}
\begin{equation}
B_{\theta'}=\frac{B_0}{2r^3}(\sin\theta'\cos\alpha+\cos\phi'\cos\theta'\sin\alpha),
\label{bt1}
\end{equation}
\begin{equation}
B_{\phi'}=\frac{B_0}{2r^3}\cos\phi'\sin\alpha.
\label{bp1}
\end{equation}
\end{mathletters}
The magnetic-field component parallel to the rotation axis, $B_{z'}$,
is
\begin{equation}
B_{z'}=B_{r'}\cos\theta'-B_{\theta'}\sin\theta'.
\label{bz1}
\end{equation}
For the point of intersection between the null charge surface and the
light cylinder, the following equations should be satisfied,
$$
B_{z'}=0,~r'\sin\theta'=\frac{cP}{2\pi},
$$
where $P$ is the period of the pulsar.
Transform these two equations into the initial magnetic coordinates,
\begin{equation}
\frac{(\sin\theta\cos\phi\cos\alpha+\cos\theta\sin\alpha)^2+\sin^2\theta\sin^2\phi-
2(-\sin\theta\cos\phi\sin\alpha+\cos\theta\cos\alpha)^2}
{3(\sin\theta\cos\phi\cos\alpha+\cos\theta\sin\alpha)
(-\sin\theta\cos\phi\sin\alpha+\cos\theta\cos\alpha)}=\tan\alpha,
\label{aeq1}
\end{equation}
\begin{equation}
r\sqrt{(\sin\theta\cos\phi\cos\alpha+\cos\theta\sin\alpha)^2+\sin^2\theta\sin^2\phi}
=\frac{cP}{2\pi}.
\label{aeq2}
\end{equation}
With Eq.~\ref{eq4}, the zenith angle of the intersection point between
the critical field line and the pulsar surface $\theta_0$ can be
calculated,
\begin{equation}
\frac{\sin^2\theta_0}{R}=\frac{\sin^2\theta}{r}.
\label{aeq3}
\end{equation}
Eq.~\ref{aeq1}-\ref{aeq3} give the relation
between $\theta_0$ and $\phi$, and the critical field line can be
determined by solving these equations.

The situation in which the field line could intersect the light cylinder
on the opposite side should also be considered.
In the critical case, the field line in the plane of the rotation and
the magnetic axes would be simultaneously tangential and perpendicular
to the light cylinder.
Considering a field line, it has a tangent line (the tangent point is
$(r_1\sin\theta_1,~r_1\cos\theta_1)$ in rectangular coordinates) that
is parallel to one of its normal lines (the footpoint is
$(r_2\sin\theta_2,~r_2\cos\theta_2)$ in rectangular coordinates), and
the distances between the pulsar center and these two lines are equal,
\begin{equation}
\frac{x-R_e\sin^3\theta_1}{y-R_e\cos\theta_1\sin^2\theta_1}=
\frac{3\sin^2\theta_1\cos\theta_1}{2\sin\theta_1\cos^2\theta_1-\sin^3\theta_1}=k,
\label{aeq4}
\end{equation}
\begin{equation}
\frac{x-R_e\sin^3\theta_2}{y-R_e\cos\theta_2\sin^2\theta_2}=
-\frac{2\sin\theta_2\cos^2\theta_2-\sin^3\theta_2}{3\sin^2\theta_2\cos\theta_2}=k,
\label{aeq4}
\end{equation}
\begin{equation}
|R_e\sin^3\theta_1-kR_e\cos\theta_1\sin^2\theta_1|=
|R_e\sin^3\theta_2-kR_e\cos\theta_2\sin^2\theta_2|=
\frac{cP}{2\pi}\sqrt{k^2+1}.
\label{aeq4}
\end{equation}
Based on these equations, $k = 2.87322$, and this means that the
critical inclination angle $\alpha_c=\arctan k=70.8\degree$.
For PSR B1133$+$16, $\alpha=51.3\degree<\alpha_c$, so this case does
not need to be considered.

\section{B.~~Simulation}

\begin{figure}[tbh]
\centering
\subfigure{
\includegraphics[width=0.5\textwidth]{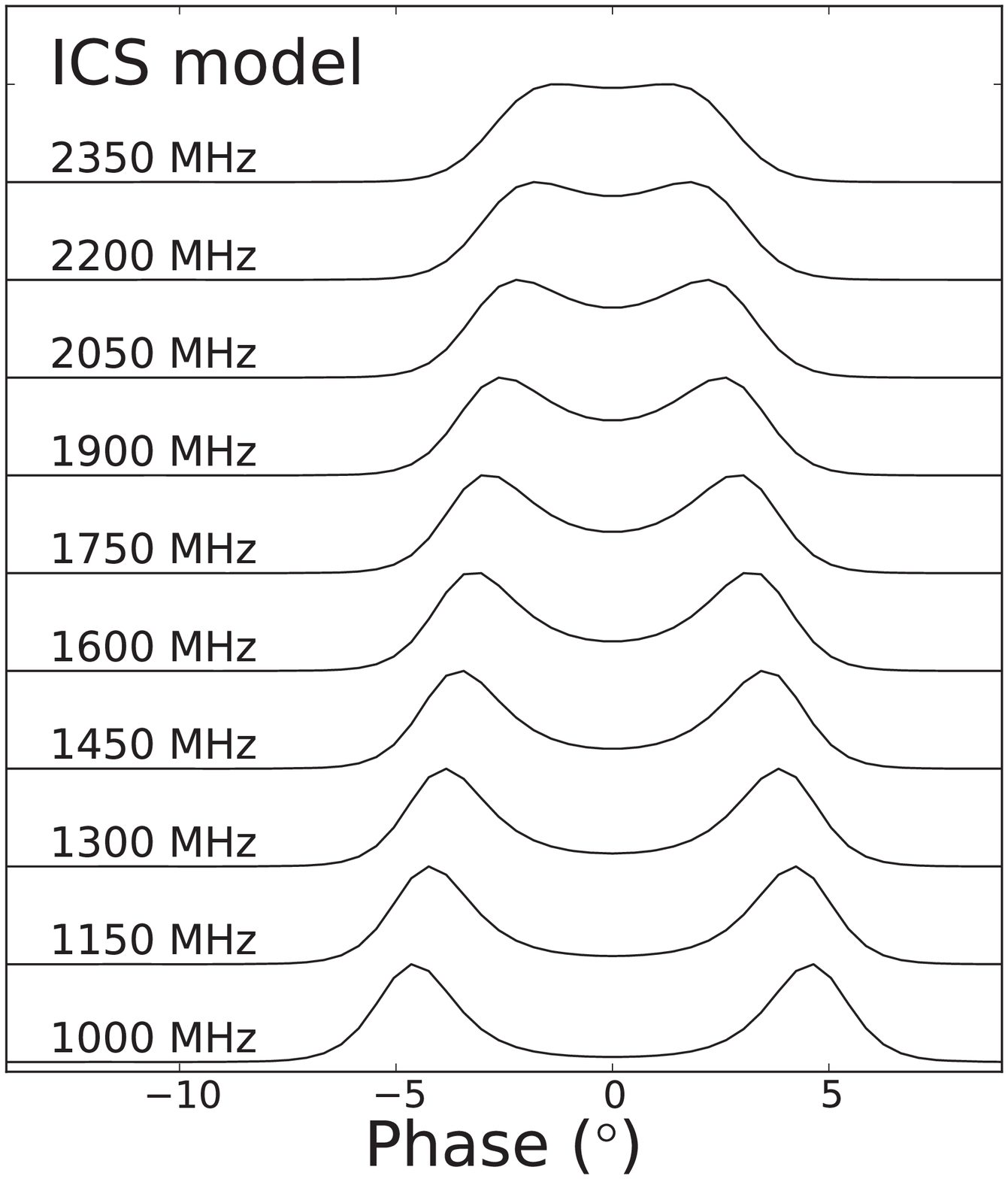}
\includegraphics[width=0.5\textwidth]{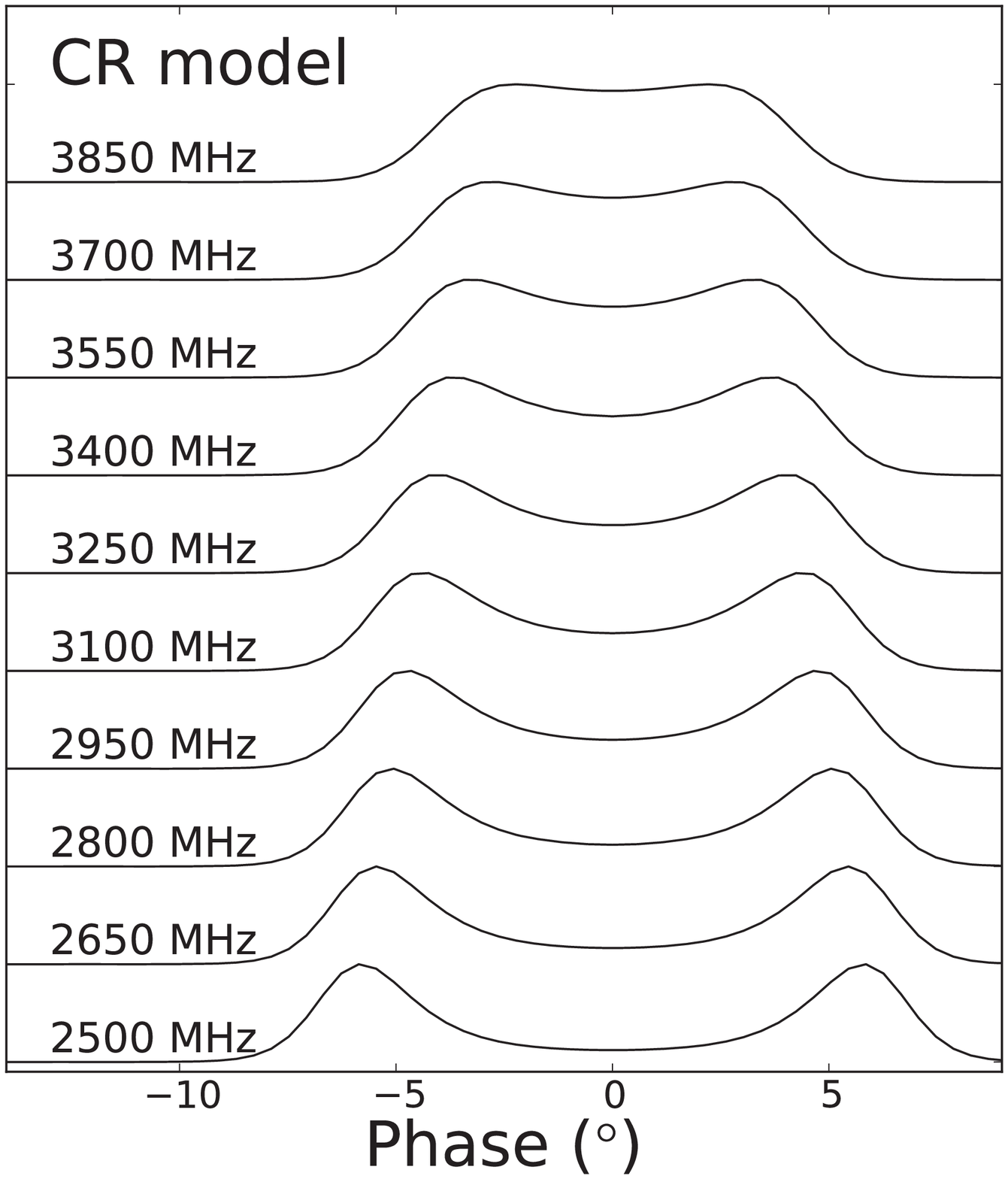}
}
\caption{Simulated multi-frequency profile of a model pulsar under the
  assumptions pertaining to the ICS and CR models.
\label{simulation}}
\end{figure}

A real profile is hard to reproduce because of the complexity
of the physical processes in the magnetosphere. Although the
evolution and distribution of particle energy have been
obtained by profiles fits, the results cover just part of the
profile around the peak center. Here a simple simulation of a
model pulsar is shown, which can reproduce some profile
properties as seen in PSR B1133$+$16.

The distribution of the particles in the magnetosphere $f$ is assumed
to follow
\begin{equation}
f=f(r,\theta,\phi,\gamma).
\label{dis}
\end{equation}
Assume that the normalized particle spectrum $s(r,\theta,\phi,\gamma)$
and the spatial particle distribution are independent of each other,
\begin{equation}
f=f_0(\theta_i,\phi)(\frac{r}{R})^{-3}\times s(r,\theta,\phi,\gamma),
\label{dis1}
\end{equation}
where $f_0$ is the initial number density function of the particles in
the polar gap, $\theta_i$ is the initial zenith angle of the field
line, and the factor $(\frac{r}{R})^{-3}$ represents the particle
dispersion while moving along the field line.
Here the assumption is that only one field line in each phase
dominates the radiation, e.g., the core-region or the annular-region
field line that we used to analyze the profile,
\begin{equation}
f_0(\theta_i,\phi)=F_0\times\delta(\theta_i-\theta_{i0}(\phi)),
\label{idis}
\end{equation}
where $F_0$ is a constant, $\theta_{i0}$ is the zenith angle of the
specified field line, and the core-region field line is adopted here.
The normalized particle spectrum is given by
\begin{equation}
s(r,\theta,\phi,\gamma)=s(\frac{\gamma-\gamma_1}{\Delta\gamma}),~~
\int^\infty_0s\mathrm{d}\gamma=1,
\label{spec}
\end{equation}
where $\gamma_1$ is the mean energy of the particles and
$\Delta\gamma$ is their energy dispersion.
Assume that the particle energy-loss rate d$(r)$ is independent of the
pulse phase,
$$
\gamma_1=\Gamma_0\times\delta(\theta_i-\theta_{i0}(\phi))\times d(r),
$$
where $\Gamma_0$ is a constant.
Using the particle distribution function and the assumption of
coherent radiation, the radiation intensity in any direction can be
calculated,
\begin{equation}
\mathrm{In~ICS~model},~I_{\mathrm{ICS}}\propto\int^\infty_R f^2(\frac{r}{R})^{-2}\gamma^2\mathrm{d}r,
\label{ics_rad}
\end{equation}
\begin{equation}
\mathrm{In~CR~model},~I_{\mathrm{CR}}\propto\int^\infty_R f^2\gamma\mathrm{d}r,
\label{cr_rad}
\end{equation}
where the factor $(\frac{r}{R})^{-2}$ represents the photons
dispersion while spreading outwards.
The simulation results are shown in Fig.~\ref{simulation}, revealing that the profile contracts while the
frequency increases. Of course, the simulated evolution of the
profile with frequency is much faster than that in real data. In
fact, this just depends on the functions $s$ and $d$,
and the degree
of coherence, which are all uncertain. Since the radiation phase
could be transformed to the radiation altitude in the magnetosphere,
the phase¨Cfrequency mapping of the pulse profile could
correspond one-to-one to the radiation spectrum of different
points on the magnetic field line at different heights.
Furthermore, the energy spectrum of particles at different height
could be obtained. In this simulation, the functions $s$ and $d$ describe the evolution of the particle spectrum with height. Thus,
the simulated profile evolution can of course always be roughly
adjusted to be similar to that of the real data as appropriate
functions and degree of coherence are used. In fact, any
centrosymmetric profile could be obtained. Then, in Fig.~\ref{simulation}, only two exemplary results are shown, which show different
speeds of profile evolution in different frequency bands.

\end{document}